\documentclass[11pt,journal,onecolumn]{IEEEtran} 
\usepackage[left=1.25in,right=1.25in,top=1in,bottom=1in]{geometry}
\usepackage{setspace}
\usepackage{amsmath,amssymb,amsthm}
\usepackage{graphicx}
\usepackage[style=numeric,bibstyle=ieee,maxbibnames=999]{biblatex}
\begin{document}
\title{Physics-driven Deep Learning for PET/MRI}
\author{Abhejit~Rajagopal$^1$, Andrew~P.~Leynes$^{1,2}$, Nicholas~Dwork$^{1,4}$, Jessica~E.~Scholey$^3$,\\ Thomas~A.~Hope$^1$, and Peder~E.Z.~Larson$^{1,2}$
\thanks{$^1$Department of Radiology and Biomedical Imaging, University of California, San Francisco, USA 94158}
\thanks{$^2$UC Berkeley-UCSF Joint Graduate Program in Bioengineering, University of California, San Francisco and University of California, Berkeley, California}
\thanks{$^3$Department of Radiation Oncology, University of California, San Francisco, USA 94158}
\thanks{$^4$Departments of Pediatrics and Radiology, University of Colorado, Anschutz, USA 80045}
\thanks{Preprint, under review.}}
\maketitle
\vspace{-10mm}
\begin{abstract}
In this paper, we review physics- and data-driven reconstruction techniques for simultaneous positron emission tomography (PET) / magnetic resonance imaging (MRI) systems, which have significant advantages for clinical imaging of cancer, neurological disorders, and heart disease. These reconstruction approaches utilize priors, either structural or statistical, together with a physics-based description of the PET system response. However, due to the nested representation of the forward problem, direct PET/MRI reconstruction is a nonlinear problem. We elucidate how a multi-faceted approach accommodates hybrid data- and physics-driven machine learning for reconstruction of 3D PET/MRI, summarizing important deep learning developments made in the last 5 years to address attenuation correction, scattering, low photon counts, and data consistency.  We also describe how applications of these multi-modality approaches extend beyond PET/MRI to improving accuracy in radiation therapy planning. We conclude by discussing opportunities for extending the current state-of-the-art following the latest trends in physics- and deep learning-based computational imaging and next-generation detector hardware.
\end{abstract}
\IEEEpeerreviewmaketitle

\vspace{-4mm}
\section{Introduction}
Simultaneous positron emission tomography and magnetic resonance imaging (PET/MRI) systems are appealing because they combine the excellent anatomical and structural contrast of MRI with information about tissue metabolism from PET radiotracers~\cite{ehman2017}.

Historically, PET was acquired on PET/CT systems, where CT directly provides attenuation maps used to accurately weight the projection data and perform image reconstruction.  However, PET/MRI is beneficial for imaging many anatomical regions (particularly the head and pelvis) because it provides significant soft tissue contrast between neighboring anatomical structures that are difficult to distinguish with CT.
Clinical applications include diagnosing and treating patients with epilepsy, brain tumors, head and neck tumors, and cancers of the rectum, cervix, endometrium, ovaries, and prostate, and is often superior to what can be achieved by registering PET/CT with a separate MRI acquisition~\cite{zaidi2016promise}. PET/MRI is also safer for pediatric populations because it saves patients from the additional dose of ionizing radiation required for CT, which increases the lifetime cancer risk.

PET/MRI additionally presents opportunities to improve PET reconstructions by incorporating high-resolution, high-contrast, and dynamic MRI data. This has been attempted using fully-convolutional neural network (CNN) image-to-image architectures applied directly to PET/MRI reconstruction for the enhancement of noisy low-dose PET/MRI via supervised and adversarial objectives~\cite{dong2020deep,yang2020ct}, and recently for one-shot (single-stage) reconstruction via the unsupervised deep image prior technique~\cite{gong2018pet,gong2021direct}. Deep learning-based domain translation from MRI to CT has also shown improved PET/MRI quantification when combined with traditional image reconstruction~\cite{leynes2018synthetic}, especially in scenarios with data mismatch~\cite{leynes2020estimation}, and has received substantial attention over the last decade in the field of radiotherapy for MRI-only treatment planning~\cite{boulanger2021deep}. 

Despite the success of such single-stage data-driven deep learning PET/MRI reconstruction techniques in the literature, clinical adoption of these techniques remains low, perhaps due to concerns of model generalization~\cite{antun2020instabilities}. In contrast, hybrid physics-driven deep learning has quickly established prominence in PET/MRI due to its increased robustness, interpretability, and data consistency guarantees~\cite{yokota2019dynamic,leynes2020estimation}. This is evidenced by recent vendor-supplied hybrid deep learning reconstruction algorithms. While these primarily utilize deep learning-generated MR-based attenuation correction, recent work has also addressed innovations to the PET reconstruction algorithm by directly incorporating deep learning-derived priors (e.g., joint total-variation or joint-sparsity)~\cite{rajagopal2021enhanced}.

In this article, we will provide a thorough overview of these developments and provide context for how they ultimately improve the resolution and signal-to-noise ratio of 3D PET/MRI, especially for clinical scanning of human subjects.
Although there have been many IEEE publications about deep learning and physics-based reconstructions, the application to PET/MRI reconstruction is particularly interesting because of the multi-faceted, non-linear structure of the problem.  Robust solutions incorporate both PET and MRI domain knowledge, including the physics of radioactivity, electromagnetics, and detector hardware, as well as the expected biological distributions and their variations between PET and MRI.  Thus we believe presenting physics-based deep learning for simultaneous PET/MRI will also have broad interest for other multi-modality imaging applications.

\subsection{Challenges and Opportunities}
The major challenge that is currently being addressed by physics-based deep learning in PET/MRI is to determine the attenuation of the 511-keV photons used in PET as they travel through the body.  This is a critical component of PET reconstruction, as there are large differences in attenuation across the field-of-view, particularly between air (low attenuation), soft-tissue (moderate attenuation), and bone (very high attenuation).  

In PET/CT, the CT intensity (in Hounsfield Units) strongly correlates with the electron density; none of the MRI contrasts (proton density,  T1, T2, chemical shift, diffusion, flow, magnetic susceptibility, etc) can be used to generate photon attenuation maps directly.
Moreover, with conventional MRI pulse sequences, there is negligible signal from bone, which is critical to detect since it is the most strongly attenuating tissue type. Thus, a major limitation for accurate PET/MRI is the prediction and use of accurate attenuation maps.

The major opportunities that can be addressed by physics-based deep learning in simultaneous PET/MRI are to improve the PET resolution and signal-to-noise ratio (SNR), particularly in the presence of motion, short scan times, and low radionuclide dose.
This presents a substantial opportunity for PET/MRI to improve the PET reconstruction by utilizing both simultaneous and historical MRI data. The excellent tissue contrast of MRI means that most structures and boundaries can be measured on MRI with higher spatial resolution compared to PET and increased soft-tissue contrast compared to CT. MRI can be used directly as a prior for iterative PET reconstruction, and recently this approach has been improved through the use of \textit{data-driven} priors generated using neural networks. This is a natural place to incorporate deep learning with greater flexibility (e.g.~not just attenuation correction) without sacrificing the physical consistency. Another opportunity is in estimating attenuation in patients with implants, which is challenging even with CT, where ``joint-estimation'' techniques that simultaneously reconstruct PET activity and attenuation maps can be improved using MR priors. Furthermore, MRI can obtain more accurate information about motion of the subject, which can be used for motion correction of PET data. 

\section{Classical PET/MRI Reconstruction Models} \label{sec:classical} 
Positron emission tomography is an inverse problem that involves recovering the concentration of radiation (becquerel/mm$^3$) inside a volume from keV photon absorption events collected at the periphery. In the standard case, the known variables are the measured count ($y \in \mathbb{Z}^{M}$) and time ($t \in \mathbb{Z}^{M}$) of photon collection in each detector, and the forward operator $\mathcal{A}$ relating world-voxels ($p_v \in \mathbb{R}^{N \times 3}$) to detector-pixels ($p_d \in \mathbb{R}^{M \times 6}$).  The system is represented by $y = \mathcal{A} x$, where $x \in \mathbb{Z}^N$ is an unknown representing the rate of photons being emitted from each of $N$ world voxels, and $\mathcal{A}=\mathcal{A}(p_d,p_v,t,\mu)$ summarizes the probability of detecting a photon from voxel $p_v$ in detector $p_d$ at time $t$ given the geometry of the detector system and the physics of positron recombination and photon transmission via dielectric media represented by transmission map $\mu \in \mathbb{R}^N$. In the simplest case, the problem is relaxed and $\mathcal{A}$ is assumed to be a static linear operator $A(p_d,p_v,\mu): \mathbb{R}^N \rightarrow \mathbb{R}^M$, resulting in a linear problem to recover the 3D radiotracer distribution $x$ from the measured projection or sinogram data $y$ collected at the periphery\footnote{Note that we are intentionally abstracting the impact of numerous hardware advances, such as time-of-flight (ToF)
and fully 3D lines-of-response
that complement the algorithms discussed here for the localization of radioactivity.}.

This simplified linear problem is classically ill-posed due to the dearth of measurements at desirable resolutions ($N >> M$) and limited integration times in clinical scanning. This has prompted numerous algorithmic advances, including improved modeling of $\mathcal{A}$ via its point-spread-function (PSF) \cite{rapisarda2010image}, regularized objectives based on total variation and compressed sensing, and approaches utilizing machine learning to perform reconstruction~\cite{wang2014pet}. One way to connect these approaches is by examining the objective function they seek to minimize.  Consider
\begin{align}
    \underset{x}{\text{minimize}} \hspace{0.5em} \|APx-y\|_p + R(APx, x, y)
    \label{eq:classical1}
\end{align}
where the first term represents $p$-norm data-consistency between measured and simulated sinogram-projection data and $R$ represents additional regularization or penalization techniques. For example, $R(APx,x,y)=R(x)$ represents priors on the image distribution of $x$ such as in total variation ($R(x) = \| \nabla x \|_2$), BSREM ($R(x)=\sum_{j=1}^{n_v}\sum_{k=N_j} w_j w_k \frac{(x_j-x_k)^2}{x_j + x_k + \gamma |x_j - x_k|}$)~\cite{lantos2018standard}, or compressed sensing based on wavelets ($R(x) = \| \psi x \|_1$)~\cite{dwork2021utilizing}. Alternatively, these approaches can often be directly combined with compressed representations, e.g.~by replacing $x$ with a low-rank factorization $x=LR^T$, multi-scale low-rank factorization \cite{ong2016beyond}, or alternative basis decomposition~\cite{wang2014pet}. These approaches all seek to reduce the non-uniqueness of the inverse problem by constraining the space of possible solutions via expected properties or structure of radiotracer image $x$.

Despite these strides, an acute practical challenge is in accurately estimating $A$ during a PET examination, since its entries depend on run-time constants, such as the precise positions of the detectors with respect to the patient ($p_d$ given $p_v$) and the photon map ($\mu \in \mathbb{R}^N$). Thus, the multi-modality PET/MRI reconstruction problem is actually a non-linear problem of finding $\mu$ (attenuation) and possibly $p_v(t)$ (if patient motion exists) to populate $A(p_v(t), p_d, \mu)$ and then recover the radiotracer distribution $x$. The most prominent methods to address these issues include maximum likelihood estimation of attenuation and activity (MLAA) for static exams \cite{rezaei2012simultaneous}, direct estimation of attenuation from non-attenuation corrected (NAC) PET images \cite{berker2016attenuation}, as well as numerous approaches that utilize MRI for estimation of motion and attenuation~\cite{leynes2017hybrid}. We can again connect these approaches by the objectives they minimize, e.g.:
\begin{align}
    \underset{x,\, \mu,\, p_v(t)}{\text{minimize}} \hspace{0.5em} \| A(\mu,p_v(t))x - y \|_p + R(x,\mu,p_v(t))
    \label{eq:classical2}
\end{align}
where we have ignored the dependence of $\mu$ on the patient motion $p_v(t)$, and $R$ captures a large number of penalization approaches. For example, when $R=R(x)$ the same regularization techniques described above can be employed. In TOF-MLAA~\cite{rezaei2012simultaneous,mehranian2017mr}, the problem is interpreted statistically and recast as a maximization of the log-likelihood of $\lambda$ (activity) and $\mu$ (511-keV attenuation) as: $\hat{\lambda}, \hat{\mu} = \text{argmax}_{\lambda, \mu} \sum_{i=1}^M \sum_{t=1}^T \log p(y_{it}|\lambda, \mu) + \beta R(\lambda) + \gamma X(\mu)$, where the probability distribution $p(y_{it}|\lambda,\mu)$ of the measured data value in the $i$th line of response (LOR) and the $t$th TOF bin is modeled using a Poisson distribution, and $R(\lambda)$ and $X(\mu)$ are penalty functions representing priors, such as local congruence penalties (defined in neighborhoods) for activity and spatially constrained log-Gaussian penalty for attenuation.

While these techniques stand as the de-facto standard for current vendor-provided software, there are numerous issues that encourage additional development. First, there are notable examples of algorithms such as maximum likelihood expectation maximization (MLEM) that result in non-ideal source distributions upon convergence~\cite{defrise2005image},
necessitating the use of regularized reconstructions in clinical scanning. Second, classical regularized reconstructions tend to be sensitive to the regularization parameter, which can depend on subtle changes in system hardware and patient characteristics that may be hard to model or calibrate at exam time~\cite{lantos2018standard}. These issues are exacerbated in settings with limited photon counts such as low-dose PET imaging, when the selected radiotracer has a long half-life, or when the PET detector system has low sensitivity, since there is an inherent ambiguity arising from the lack of statistical confidence.

An obvious limitation of these classical approaches is that they do not directly utilize knowledge from previous exams. On the contrary, data-driven deep learning methods take advantage of historical data and can provide guidance to reduce ambiguity in the mapping from MRI amplitudes to attenuation coefficients (e.g.~multi-modality correlation) or directly in the PET reconstruction (e.g.~via structural priors). However, purely data-driven approaches lack guarantees of the validity of the reconstruction, which limits their use in medical applications. To address this, the recent development of hybrid physics-driven deep learning offers the potential to combine the benefits of classical system modeling and image reconstruction techniques together with deep learning and statistical methods based on historical data. Today there exist two major paradigms for physics-driven deep learning for PET/MRI reconstruction:~(1)~deep learning methods (both supervised and unsupervised) that incorporate physics-based constraints or penalization, and (2)~physics-driven methods that incorporate data-driven constraints or priors. In the following sections, we summarize various realizations of these techniques in the context of classical physics-based reconstruction, and identify promising avenues for future study.

\begin{figure}[hbt!]
    \centering
    \includegraphics[width=0.9\linewidth]{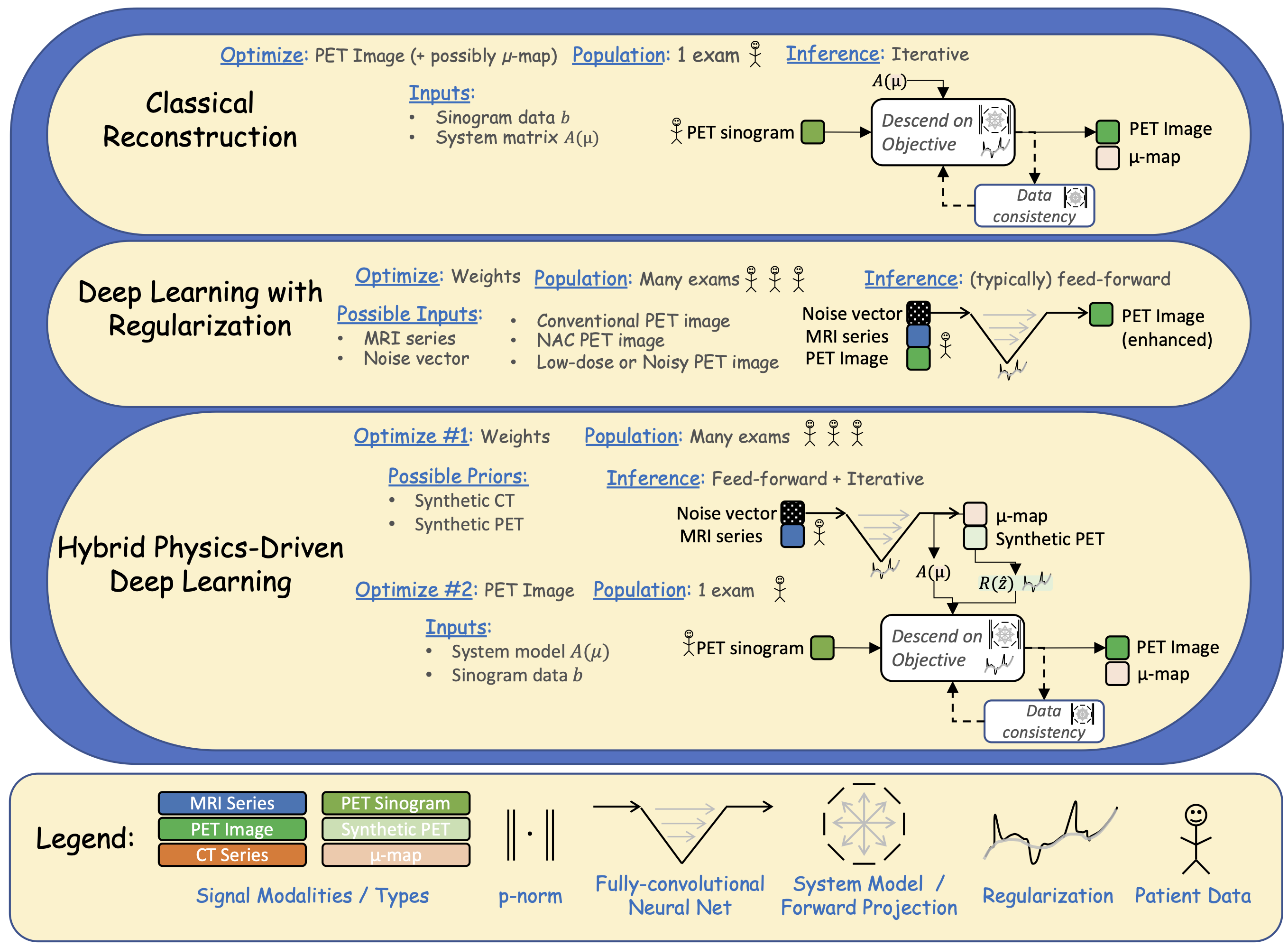}
    \vspace{-4mm}
    \caption{Comparison of classical and deep learning-based PET/MRI reconstruction algorithms, with iconography to indicate where components such as data-consistency $p$-norms, the system model, regularization, or neural networks are used. A key feature of hybrid physics-driven deep learning approaches is the combination of offline optimization on a population of exams together with the online run-time optimization to reconstruct the PET image of a single patient.}
    \label{fig:taxonomy}
\end{figure}

\section{Deep Learning Architectures with Physics-based Penalization} \label{sec:datadriven}
In addition to combating resolution limitations of classical PET reconstruction, deep learning has been proposed for PET/MRI reconstruction to overcome practical difficulties with estimation of attenuation factors, correcting scatter, and accounting for other nonlinearities in model specification that affect overall SNR and quantitation of PET. In this respect, image-domain CNNs have been extremely successful at restoring PET images without explicit modeling of these non-idealities, which otherwise require solving a non-linear problem for each exam with careful calibration and manual parameter selection. Deep learning approaches utilize a priori offline optimization of algorithm weights (or meta optimization of network structure) and apply this optimal tuning to reconstruction of subsequent exams.

\subsubsection{Direct Prediction of Attenuation-Corrected PET} A conceptually-straightforward approach is to use a deep network to predict an attenuation corrected PET image from a non-attenuation corrected (NAC) PET image with or without an MRI reference~\cite{dong2020deep,yang2020ct}. In this paradigm, model training is performed by feeding a NAC-PET image and (optionally) an MRI image (typically T1- or T2-weighted) as registered channels of an input tensor to a fully-convolutional network (e.g.~UNet) whose output is then compared with an ideal, attenuation-corrected (AC) PET image, and minimized. The ideal groundtruth PET image is typically derived from an alternative system, such as PET/CT, where there is high confidence in the conventional AC-PET reconstruction. The most common objectives for training these models are $p$-norms with regularization:
\begin{align}
    \underset{\omega}{\text{minimize}} \| f_\omega(X) - Y \|_p + R(f,f(X),Y)
    \label{eq:deep_supervised}
\end{align}
where $f_\omega:\mathbb{R}^{K\times N} \rightarrow \mathbb{R}^{N}$ represents a deep neural network with parameters $\omega$, $X \in \mathbb{R}^{K \times N}$ represents the input tensor (e.g.~$K=1$ for NAC-PET input, $K=2$ for NAC-PET and MRI input), $Y \in \mathbb{R}^N$ represents the ideal AC-PET tensor, and $R$ is a regularization function that encourages expected properties in the estimate. For example, $R=R(f)$ may be chosen to regularize the condition number of the transforms inside $f$ by controlling the magnitude of its weights $\omega$: $R(f)=\gamma \|f\|_p = \gamma \sum_k \|\omega_k\|_p$; here, $\gamma\geq0$ is the regularization parameter.
When $R=R(f(X))$, penalties applicable to Eq.~\ref{eq:classical1} may be used, although this is uncommon in neural network training due to the first term in Eq.~\ref{eq:deep_supervised} already representing image-domain voxel-wise consistency. Instead, consistency can be evaluated in a transform domain, e.g.~the natural sinogram domain when $R=R(f(X),Y)=\| A f(X) - AY \|$, where $A$ represents the system matrix, and $AY$ represents an integration along the PET lines of response. Such tomographic-type regularization has been shown to improve the quantitation of DNN-derived 3D PET~\cite{shi2019novel,rajagopal2022synthetic}.

Another popular approach is to utilize perceptual losses. Unlike for natural RGB-images, there is no consensus of a standard CNN model applicable to the interpretation of DNN-derived PET image features (e.g.~perceptual losses defined by the outputs of a pre-trained VGG-16). Instead, practioners have utilized the principles of generative adversarial networks (GANs) to enable dramatic increases in perceptual image quality, which correlate with quantitative improvements. In this paradigm, the input tensor $X$ (MRI and/or NAC-PET image) is interpreted as a seed for the generative deep network $f$, and a separate discriminator module $g: \mathbb{R}^N \rightarrow \mathbb{R}$ is trained to detect whether or not $\hat{Y}=f(X)$ belongs to the same distribution as $Y$, with the assumption that any $Y \in \{Y_k\}$ of training data is distinct from any $\hat{Y}$ produced by the generator. Thus, in relation to Eq.~\ref{eq:deep_supervised}, $R=R(f(X),Y)$ represents loss derived from training of the discriminator module $g$ using balanced batches of $\hat{Y}$ and $Y$, although in practice an alternating minimization is typically employed. The application of GANs to PET/MRI requires a strong anatomic correspondence between the seed and result images, preventing adoption of approaches that rely solely on the perceptual GAN loss without data consistency checks.

The quantitative benefit of these supervision techniques can be evaluated using metrics such mean absolute error (MAE) $\frac{1}{N} \|X_i-Y_i\|_1$, relative mean-square error (rMSE) $\frac{1}{N} \sum_i^N (X_i-Y_i)^2 / ( \gamma + Y_i )$, and structural similarity index measure~(SSIM), ${\frac {(2\mu _{X}\mu _{Y}+c_{1})(2\sigma _{XY}+c_{2})}{(\mu _{X}^{2}+\mu _{Y}^{2}+c_{1})(\sigma _{X}^{2}+\sigma _{Y}^{2}+c_{2})}}$, where $(\mu_z,\sigma_z)$ represent the mean and variance or covariances of the argument and $c_*$ is chosen proportional to the dynamic range of pixel values within a image patch~\cite{wang2003multiscale}. The issue of dynamic range is particularly important for PET image reconstruction, where the standardized uptake values (SUV) can cover a large range (e.g.~0 - 50) depending on the radiotracer and dose delivered to the patient. This is an issue that also affects the training of deep networks, since $p$-norms by themselves are typically insufficient to balance the contribution from different organ systems, requiring manual balancing using regional- or distribution-based balancing of the loss~\cite{rajagopal2022synthetic}. In this respect, SSIM is desirable for quantifying image similarity because it accounts for regional-changes in expected amplitude, resulting in a better measure of perceptual image quality matching radiologists' interpretation.

\subsubsection{Supervised Prediction of Full-dose PET from Low-dose PET} The innovations that succeeded in generating AC-PET from NAC-PET using a deep network without explicit modeling of the µ-map can be extended to other non-idealities as well, such as the prediction of ``full-dose'' PET from low-dose PET, which is typically noisier and offers less contrast for radiologist interpretation. Here, low-dose refers to a small photon count resulting from administration of a low concentration radionuclide solution or from a radionuclide with a long half-life. Since 2017, there have been a number of research works demonstrating the ability to generate high-quality PET images from low-dose PET images for a number of different radiotracers, such as 18F-florbetaben amyloid~\cite{chen2019ultra,ouyang2019ultra} and 18F-fludeoxyglucose~\cite{xu2020ultra},
using UNets and unrolled architectures. The primary goal of these works is to improve the visual interpretability of PET, as measured by MAE, rMSE, SSIM, or (in some cases) validated by a radiologist~\cite{zaharchuk2020ai}. Conceptually, the application to low-dose enhancement does not necessitate architectural changes from the scatter and attenuation-correcting networks previously described. Instead, the focus is on improving perceptual image quality and SNR given a conventional low-dose reconstruction, which can differ in both fidelity and content/structure (i.e.~spatial distribution of the radiotracer). The key requirement for this application is the availability of a large dataset of high quality full-dose PET reconstructions that can be used to train the neural network using objectives of the form in Eq.~\ref{eq:deep_supervised}. While it is possible to generate such training data for radiotracers with short half-lives by retrospectively integrating the recorded counts from full-dose exams over a shorter time window, this presents a challenge for radiotracers with long half-lives due to availability of a high-quality exemplars that take hours to collect and would thereby be corrupted by artifacts arising from both patient motion and changes in the spatial distribution of the radiotracer over the integration time period.

\subsubsection{Unsupervised Denoising Techniques based on Deep Image Prior}
In 2018, it was demonstrated that the denoising of low-dose PET can be achieved in an \textit{unsupervised} fashion, by exploiting the inherent spatial regularization properties of CNNs such as the UNet~\cite{gong2018pet}. Specifically, the Deep Image Prior (DIP) technique applied to PET/MRI images trains a randomly initialized UNet to predict a conventionally-reconstructed low-dose PET image from a input seed (e.g.~noise vector or multi-contrast MRI) for every PET/MRI exam. That is, a separate network is trained for each PET/MRI reconstruction, typically utilizing objectives of the form:
\begin{align}
    \underset{\omega}{\text{minimize}} \; \text{D}_\text{KL}(b \; || \; A f_\omega(\mu|X)) \label{eq:DIP}
\end{align}
where $b$ represents noisy projection data, $X$ represents a prior such as MRI or a noisy PET image, $\text{D}_\text{KL}$ represents the KL-divergence, and $f$ represents a neural network with parameter $\omega$ taking input noise vector $\mu$. While large networks are able to easily overfit to each exam and predict the image almost exactly, the DIP hypothesis is that, during evolution of weights $\omega$ during training, a denoised version of the PET image $\hat{x} = f_\omega(\mu)$ will be generated. This reconstruction technique utilizes the CNN's structure itself as a prior, since images that it predicts must exist in the range-space of $f_\omega$ -- a subset of $\mathbb{R}^N$ that is highly dependent on the architecture of the model including the number of layers and channels, kernel widths, normalization, and skip connections.

The key benefit of the DIP approach is that it does not require high-resolution or high-SNR training data, other than for post-training validation studies. This extends the applicability of DIP to many new problem areas, including low-dose PET/MRI with long half-lived radiotracers and dynamic imaging where there may not exist groundtruth attenuation maps or PET images~\cite{yokota2019dynamic}. However, DIP techniques consequently require a special early-stopping criteria for the optimization that captures the sharpness or quality of the reconstructed PET image without a groundtruth image for comparison. Unfortunately, there is no good consensus on optimal automated stopping criteria for DIP in PET/MRI, and several works set a maximum number of iterations that is calibrated using quantitative error metrics (SNR, SSIM, rMSE) on synthetic data with groundtruth~\cite{yokota2019dynamic} or perceptual metrics on real data (reader studies)~\cite{gong2021direct}. Thus, despite many research studies, there is limited confidence that the DIP hypothesis is always valid for PET measurements, which can be degraded by a number of phenomena including limited integration times, inaccurate attenuation estimation, photon scattering, patient motion, and statistical noise.

\begin{figure}[hbt!]
    \vspace{-4mm}
    \centering
    \includegraphics[width=\linewidth]{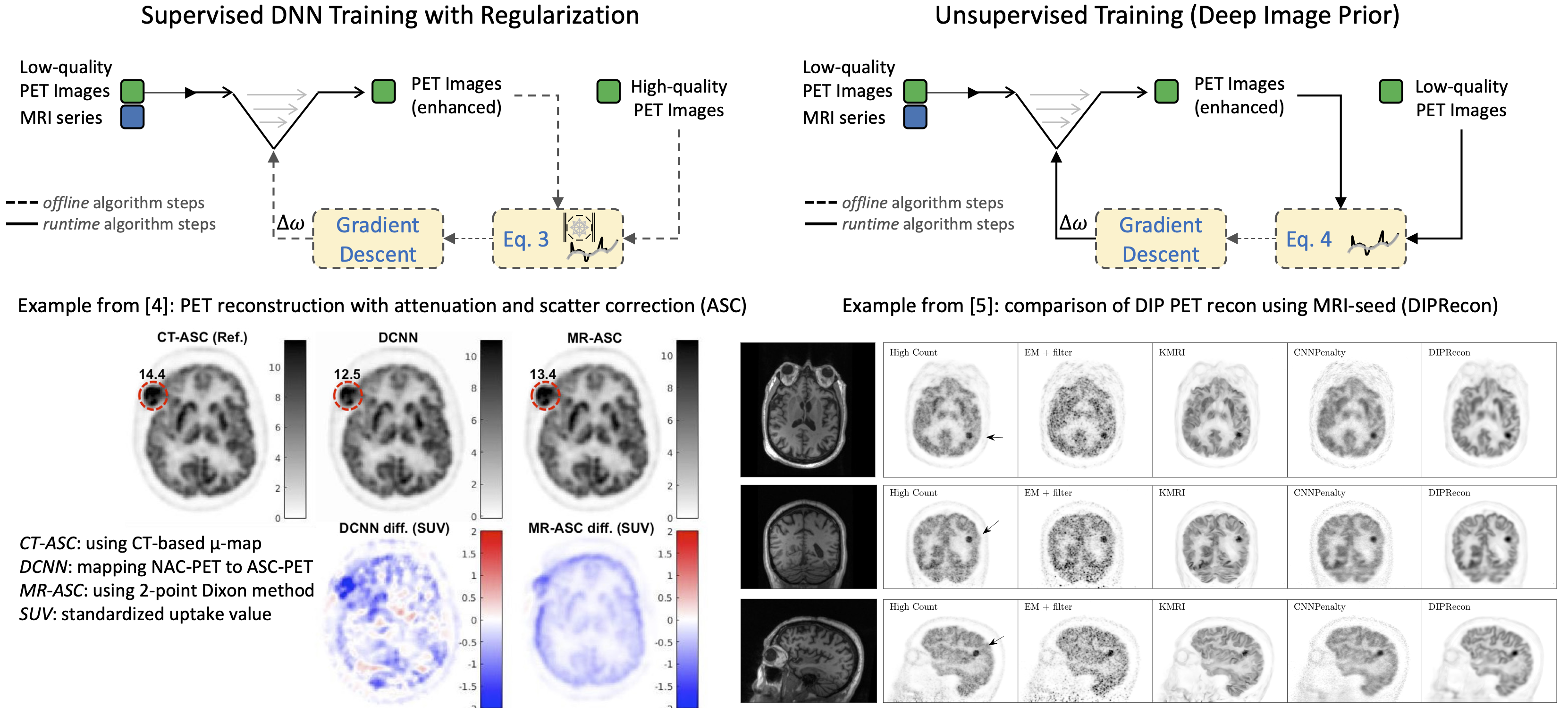}
    \vspace{-6mm}
    \caption{Schematic of training and evaluation strategies for supervised and unsupervised deep learning architectures for PET/MRI reconstruction.
    Supervised training of NAC-PET to AC-PET utilizes a offline optimization (typical objectives of the form in Eq.~\ref{eq:deep_supervised}), whereas unsupervised deep image prior (DIP) techniques utilize online optimization (typical objectives of the form Eq.~\ref{eq:DIP}). Example figures are reproduced in part from existing literature, \cite{yang2020ct} and \cite{gong2018pet} respectively.
    }
    \label{fig:supervised1}
    \vspace{-4mm}
\end{figure}

\section{Physics-Driven Methods with Data-driven Constraints} \label{sec:physicsdriven}
While it is clear that deep learning can enhance PET reconstruction, there remains a risk associated with using data-driven methods for clinical diagnosis or therapy due to the lack of strong performance guarantees. Some notable examples of this include performance degradation under dataset shift~\cite{chen2020generalization}, the existence of natural adversarial attacks, and phantom lesions. In contrast, emerging hybrid physics-driven methods for PET/MRI prioritize consistency of the reconstruction with measured projection data and use data-driven inputs as structural or statistical priors that guide the reconstruction and reduce the ill-posedness of the problem. In this section we review recent development of physics-driven methods that utilize priors derived from deep learning while offering stronger performance guarantees e.g.~due to the use of convex objectives or well-understood signal models, which increase their potential for clinical adoption.

\subsubsection{Domain Translation of MRI to CT for PET Attenuation Correction and Radiation Therapy Planning} \label{sec:MR2CT}
When gamma particles are emitted from positron recombination events, they traverse various dielectric media (tissue, blood, bone, air) before depositing energy in the detector. Although PET/MRI systems offer the excellent soft-tissue contrast of MRI for understanding and mapping anatomy, the amplitude of MRI does not readily translate to 511-keV photon attenuation. To this end, prior techniques have utilized vision- and segmentation on MRI series to generate a pseudo-CT volume~\cite{wollenweber2013comparison}, which serves as the effective photon transmission map ($\mu$) that is incorporated into the forward model of the system $A$. However, these techniques are not not flexible enough to handle diverse patient anatomies. Since 2018, fully-convolutional networks (typically variations of 2D or 3D UNets) have been used for translating single- or multi-sequence MRI to synthetic- or pseudo-CT for PET attenuation correction~\cite{leynes2018zero, torrado2019dixon}.

Due to the highly non-local and nonlinear nature of this mapping, an open question was the effect of the MR pulse sequence on pseudo-CT accuracy; with \cite{leynes2018synthetic}
determining that Dixon-type pulse sequences are sufficient for domain translation, as has been developed by \cite{torrado2019dixon}. This domain-translation capability dramatically expands the potential of PET/MRI to improve the state of the art compared to even PET/CT, such as for PET imaging in challenging scenarios involving metal implants or acyclic patient motion, which are difficult to account for in PET/CT. For example, metal typically causes scattering on CT that affects the accuracy of the attenuation map, in addition to raising safety concerns that often affect PET/CT examination protocols. MRI can be used safely as an alternative anatomical imaging modality in these scenarios, and can additionally provide time-resolved imagery that can be used for accurate motion and attenuation compensation~\cite{kolbitsch2018respiratory}. One challenge for direct domain translation, however, is the co-registration of CT and MRI pairs. Since CT and MRI cannot be acquired simultaneously, there may be positioning errors between two scans. Thus, methods based on GANs have gained traction as these do not require 1:1 correspondence at all voxel positions~\cite{gong2020mr},
although this is ultimately desirable to encourage physical correspondence between anatomy in the MRI and synthetic CT.

\begin{figure}[hbt!]
    \centering
    \includegraphics[width=1.0\textwidth]{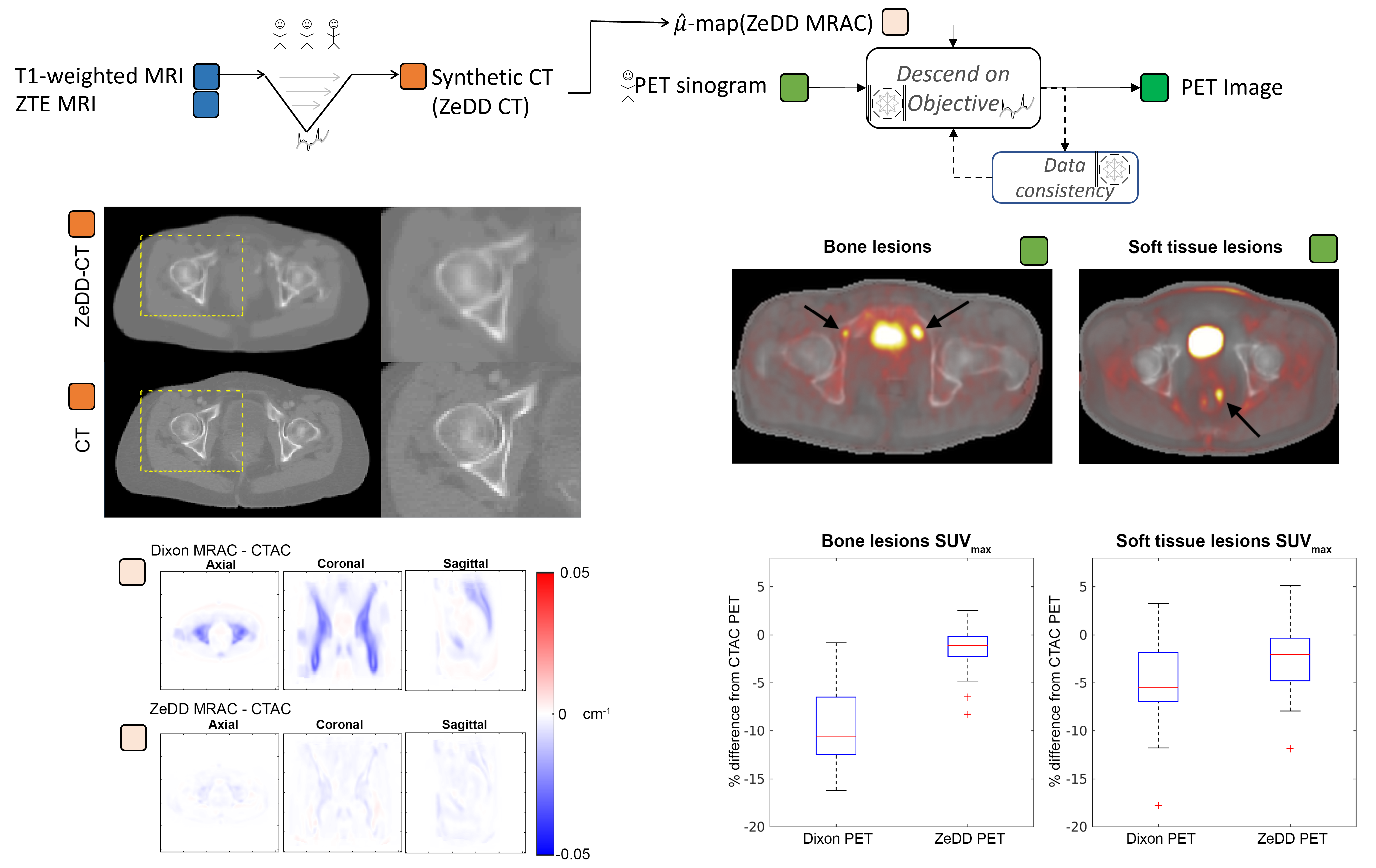}
    \vspace{-8mm}
    \caption{Effect of attenuation correction on a pelvic PET/MRI exam. The synthetic CT images produced by a deep neural network (ZeDD-CT) are similar to the real CT image albeit with a lack of sharpness (top-left). However, on the attenuation coefficient maps, there is much less error compared to the previous approach (Dixon MRAC) that is only able to estimate attenuation coefficients in soft tissues and air (bottom-left). This led to improvements in maximum standardized uptake value (SUV$_\text{max}$ estimates) in bone and soft-tissue lesions that is closer to the gold standard of PET/CT using CT attenuation correction (CTAC) (right)}
    \label{fig:my_label}
\end{figure}

A very similar problem to PET reconstruction is radiation therapy planning. With PET, one injects a solution with a radionuclide which emits gamma particles approximately along a line. The detection of these particles is used for a tomographic image reconstruction of the radionuclide or radiotracer distribution. With radiotherapy planning, the process is almost the opposite;
where one seeks to transmit radiation from an external beam with the intent of depositing a prescribed radiation dose at specified anatomic locations where--and only where--radiation treatment is necessary (e.g.~tumors).
Similar to PET/MRI image reconstruction, a challenge in radiotherapy planning is accurately estimating photon attenuation coefficients of different tissues, or effective transmission map $\mu$, to precisely deliver radiation dose (measured in units of Gray, Gy) to the intended location while avoiding radiation accumulation in healthy or sensitive areas.

CT has been the presiding imaging modality in the field of radiotherapy due to its accurate electron density mapping required for dose calculation, but MRI provides superior soft tissue contrast which is advantageous for visualizing many different tumors. Domain translation of MRI to CT provides substantial promise for an MRI-only radiotherapy workflow and has been demonstrated in a wide array of disease sites including the brain, pelvis, head-and-neck, and abdomen~\cite{owrangi2018mri}. Of these approaches, the most common include generator-only and GAN-based architectures with single-input (typically standard T1 or T2-weighted MRI) and single-output (sCT)~\cite{boulanger2021deep}. Notably, this application places new constraints on the quality and robustness of deep domain translation techniques, as erroneous mapping of MRI to CT may result in under-dosing of tumors and/or over-dosing of nearby critical healthy structures. Standardization of sCT evaluation for radiotherapy applications remains a key challenge for successful implementation of this technology.

\begin{figure}[hbt!]
    \centering
    \includegraphics[width=1.0\textwidth]{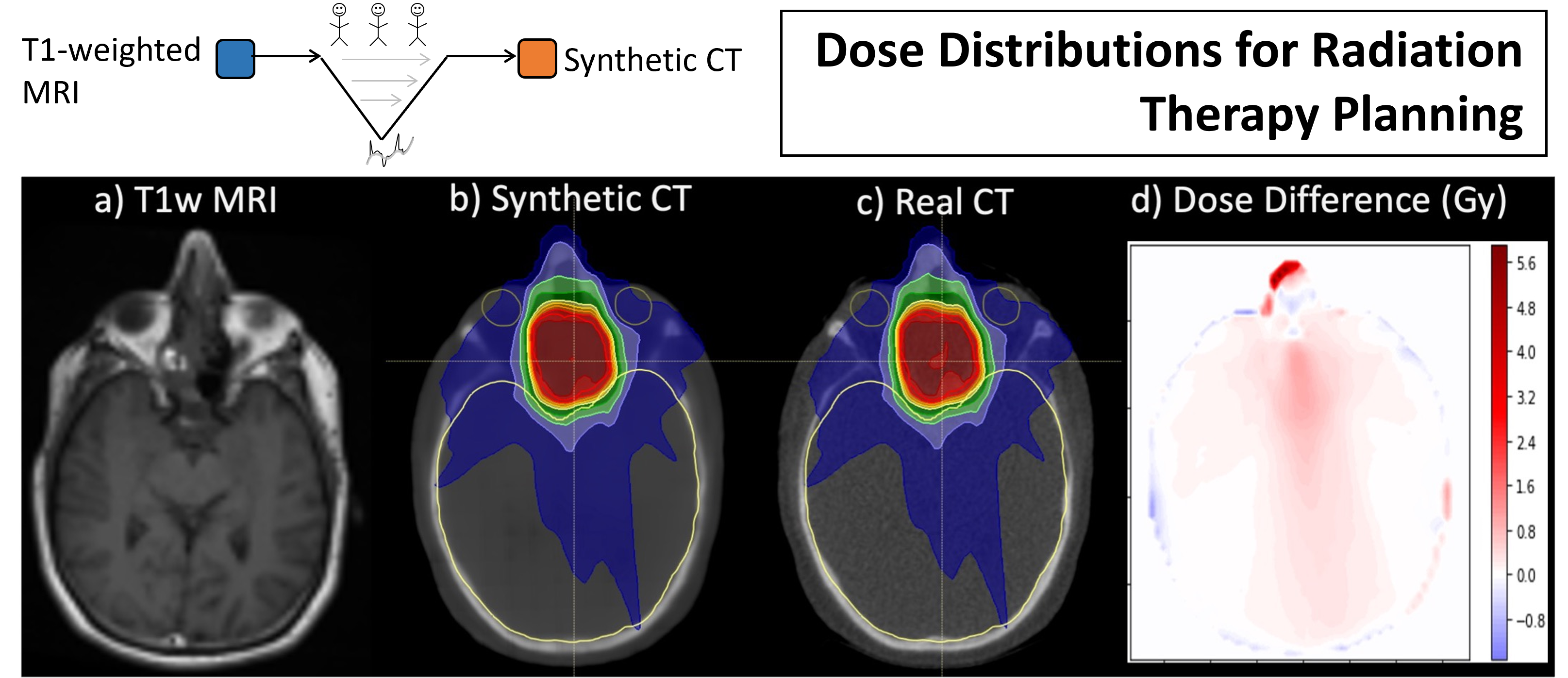}
    \vspace{-8mm}
    \caption{Exemplar image demonstrating a single-input T1-weighted MRI (a) and single-output synthetic CT (b) generated from a 3D U-Net architecture for MRI-only radiotherapy treatment planning for a patient treated to a paranasal sinus tumor (66 Gy prescription dose). The dose distribution calculated on the real CT (c) is shown for comparison, along with dose difference maps between synthetic and real CTs (d). Note some dose distribution differences at the anterior-nasal region where there is some disagreement in cartilage structure between real and synthetic CT images. }
    \label{fig:radiotherapy}
\end{figure}

\subsubsection{Joint-Reconstruction of Attenuation and Activity}
 
The joint reconstruction of attenuation and activity or the so-called maximum likelihood estimation of attenuation and activity (MLAA) is used to solve for activity images and attenuation maps in an alternating optimization scheme~\cite{rezaei2012simultaneous}. Since synthetic CT images were already being developed for PET/MRI attenuation correction, these were then incorporated as priors that constrain and regularize the attenuation map estimates in non-ToF \cite{benoit2016optimized} and ToF \cite{ahnJointEstimationActivity2018} PET/MRI to reduce cross-talk artifacts. The method by Ahn et al \cite{ahnJointEstimationActivity2018} was designed for estimation of metal implants and internal air in whole-body PET/MRI. Alternatively, several other methods use deep learning to denoise the attenuation maps estimated through MLAA \cite{hwangGenerationPETAttenuation2019} by training a model to translate the noisy attenuation map estimated by MLAA to attenuation maps derived from CT.  These methods require a high-quality training dataset to train the model. This becomes challenging in anatomical areas such as the pelvis, where there is bowel air that shifts between scan sessions, and there may be artifacts that appear differently between CT and MRI (e.g. metal artifacts or truncation artifacts)
. Leynes et al developed a Bayesian deep learning approach to generate synthetic CT images from MRI that includes estimates of uncertainty \cite{leynes2020estimation}. This uncertainty map is then used to automatically adjust for the strength of the synthetic CT priors as an extension of the work of Ahn et al \cite{ahnJointEstimationActivity2018}. The uncertainty estimates highlighted regions of bowel air mismatch, arm truncation, and metal implant artifacts with high uncertainty and emission data was used to reconstruct for these regions. There was low uncertainty in regions of normal anatomy and the synthetic CT prior was more dominant in these regions. In addition to avoiding potential convergence issues with MLAA and achieving better overall performance in normal anatomy, this combination improves the accuracy of estimated of attenuation coefficients in metal implant regions and internal air, which provide a benefit over PET/CT.

\begin{figure}[hbt!]
    \centering
   \includegraphics[width=\linewidth]{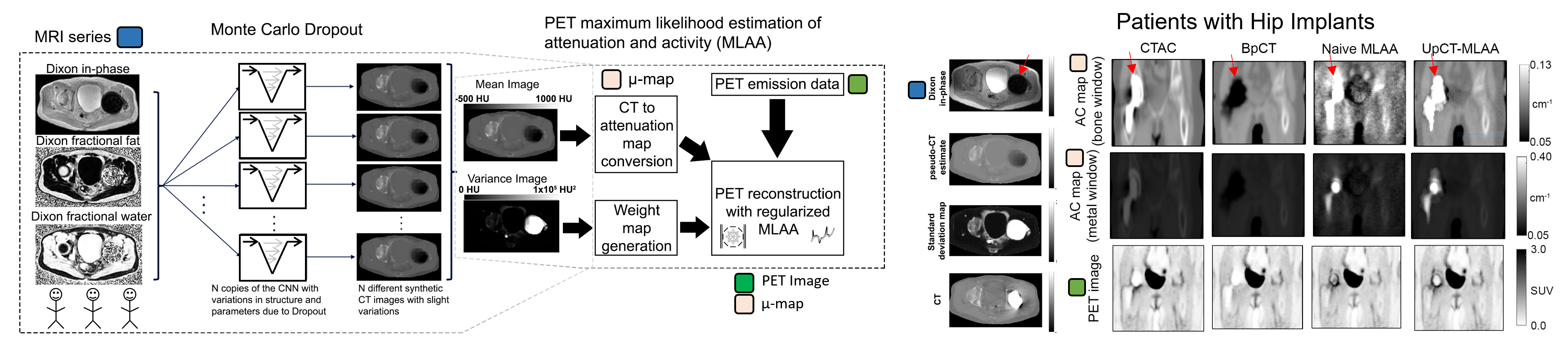}
    \caption{Incorporation of deep learned priors and uncertainty-based weighting with MLAA (UpCT-MLAA~\cite{leynes2020estimation}.  A Bayesian deep learning approach based on Monte Carlo dropout yields both synthetic CT and uncertainty priors for a MLAA PET reconstruction. In these patients with metal implants (red arrows), domain translation methods (BpCT) fail due to the presentation of the metal artifact as a signal void that corresponds to air. MLAA methods by themselves yield poor quality attenuation correction maps. The Bayesian deep learning uncertainty-driven weighting of priors in UpCT-MLAA allows for automatic weighting inside and outside the implant region that resolves both challenges.}
    \label{fig:hybridMLAA}
    \vspace{-2mm}
\end{figure}

\subsubsection{Joint-Sparsity Priors Derived from Deep Learning }
In 2020 a new hybrid approach to enhance low-dose PET reconstruction was introduced that combines a UNet-based zero-dose synthetic PET image with di-chromatic interpolation of conventionally-reconstructed low-dose PET images to yield PET images that is provably consistent with measurements but with higher SNR~\cite{rajagopal2021enhanced,dwork2021di}. Here, synthetic PET refers to PET-like images generated through domain translation of MRI, crucially without incorporation of any functional measurements of radiotracter activity (i.e.~zero-dose). Synthetc PET was originally developed for generation of anthropomorphic digital PET phantoms with realistic physiological uptake, which is achieved by training a 3D residual UNet with an objective combining the conventional $p$-norm error with a projection or line integral loss that acts as a perceptual loss for tomography~\cite{rajagopal2022synthetic}, as:
\begin{align}
    \underset{\omega}{\text{minimize}}\; || x - f_\omega(z) ||_p + \lambda || P x - P f_\omega(z) ||_p
\end{align}
where $z$ is the input MRI (e.g.~T1, T2, or Dixon water image), $x$ is the groundtruth PET image (e.g.~produced by conventional ToF-OSEM reconstruction), $f_\omega$ is the domain translating 3D UNet, and $P$ is a system/projection matrix so $Px$ and $P f_\theta(z)$ represent simulated singorams. This optimization of neural network parameters $\omega$ is performed offline using a large training dataset (paired MRI and PET reconstructions) and fixed for subsequent processing steps.

Of course, since MRI does not contain functional information about the radiotracer distribution of a patient, the synthetic PET image cannot be used directly for patient-specific analysis or diagnosis. Instead, the key idea of \cite{rajagopal2021enhanced} method is to utilize domain-translated images $\tilde{x}=f_\theta(z)$ as data-driven priors within joint-sparsity-like objective terms (e.g.~joint total-variation) of the traditional PET reconstruction problem. A performance gain is possible over conventional joint-sparsity reconstruction based on MRI because these priors can be better modality-matched to the PET tracer distribution. For example, a prior $\tilde{x}$ may be incorporated into the joint-sparsity reconstruction problem:
\begin{align}
    \underset{x}{\text{minimize}}\; ||Ax - b||_p + \lambda ||[\Psi x,  \Psi \tilde{x}]^T||_1 \label{eq:jointsparsity2}
\end{align}
where $x$ represents the PET image to be found, $A$ represents the system matrix or forward operator, $b$ represents the measured sinogram or projection data, $\psi$ represents a sparsifying transform (e.g.~wavelet transform), and we pick $\tilde{x}=f_\theta(z)$ as the synthetic PET image.

Another approach to incorporate structure, which can avoid the need to repeat the reconstruction process using the system matrix, is to utilize recent developments in multi-modality interpolation. Dwork et al.'s recent work~\cite{dwork2021di} on dichromatic interpolation, for example, can be used to upsample a low-resolution image (e.g.~downsampled to mitigate noise) using PET-domain-translated MRI, $\tilde{x}$, as a structural and statistically-representative template. The objective function in this case is:
\begin{equation}
  \begin{aligned}
    \underset{x\in [0,1]}{\text{minimize}} &\hspace{8pt} \| D\,B \, x - x_0 \|_{Fr}^2 +
      \lambda \left\| \nabla x - \nabla \tilde{x} \right\|_{w,2}^2
  \end{aligned}
  \label{eq:superRes}
\end{equation}
where $x_0$ is the initial (conventionally-reconstructed) count-limited PET image, $\tilde{x}$ is the domain-translated synthetic PET image, $x$ is the dichromatically interpolated/enhanced PET image, $B$ is taken to be a blurring operator, $D$ is a downsampling operator, $w=1$, and $\lambda$ is user-provided a Tikhonov-style regularization parameter. The key idea here is that downsampling noisy PET imagery increases SNR at the expense of reducing resolution and contrast, but this can be recovered by borrowing gradient information from the synthetic PET image.

Notably, by utilizing convex optimization codes that guarantee the data consistency term is below a given threshold, this approach guarantees consistency in the measurements with respect to the signal equation used (based on either the forward model $A$ or downsampling $DB$), with no failure cases observed in \cite{rajagopal2021enhanced} for example. Examples of SNR and SSIM improvements on slices of retrospectively resampled whole-body PET imagery are shown in Figure~\ref{fig:hybridJointSparsity} below.

\begin{figure}[hbt!]
    \centering
    \vspace{-4mm}
    \includegraphics[height=5.5in]{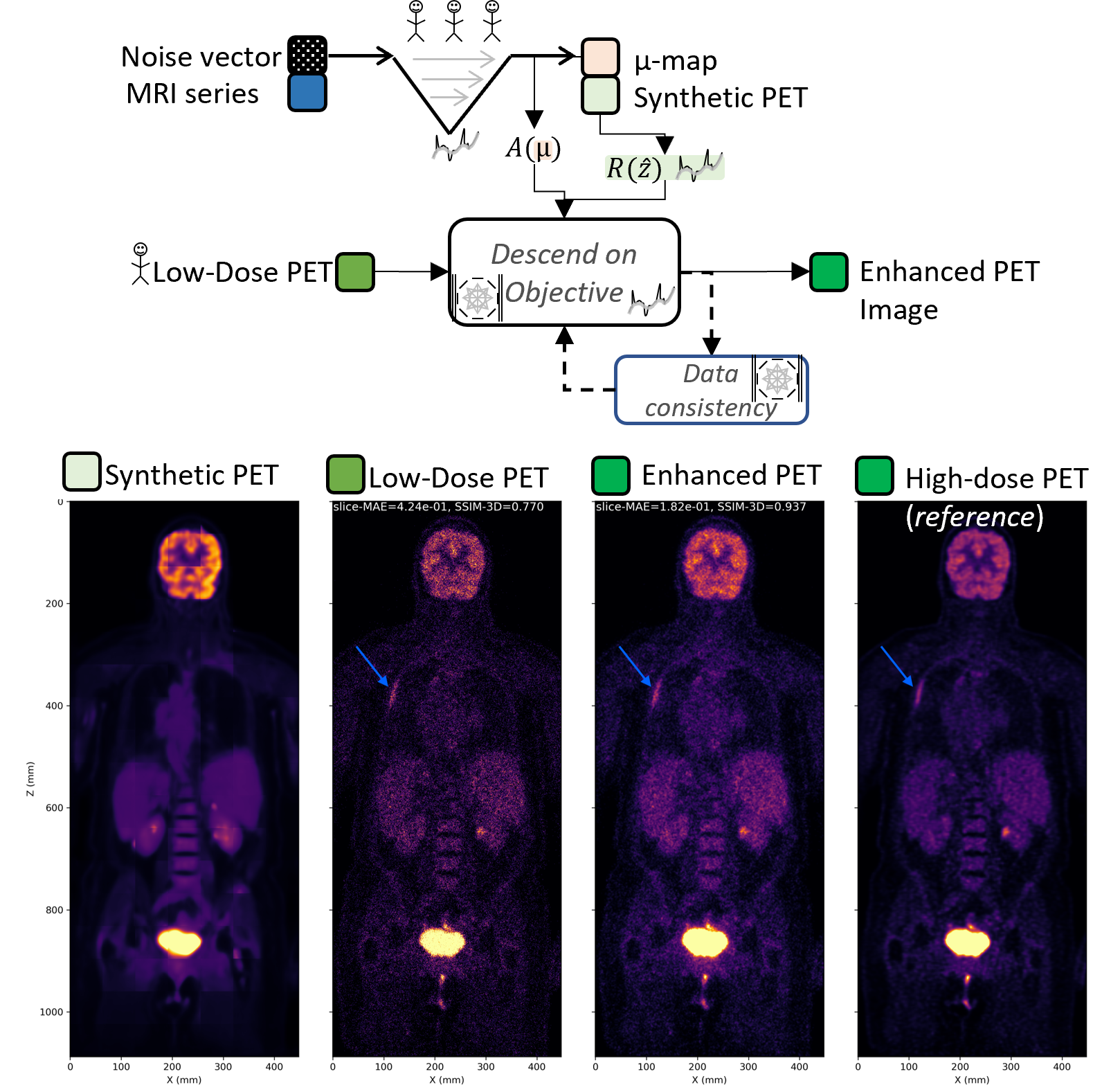}
    \vspace{-4mm}
    \caption{Example of low-dose PET reconstructions enhanced through a joint-sparsity prior based on MR-derived synthetic PET from deep learning~\cite{rajagopal2021enhanced}. Transverse slices of the 3D volumes corresponding to MR-derived synthetic FDG-PET, simulated low-dose $^{18}$F-FDG PET, dichromatically-interpolated FDG-PET, and full-dose $^{18}$F-FDG-PET, shown with scale of [0,6] SUV, illustrate that the dichromatic interpolation achieves performance from low-dose PET data that is comparable to a full-dose PET reconstruction.}
    \label{fig:hybridJointSparsity}
\end{figure}

\vspace{-6mm}
\section{Conclusion}
\subsection{Current State of the Art}
Deep learning PET/MRI reconstruction is just now becoming available directly from system vendors (e.g.~GE, Siemens, Phillips, United Imaging), which is a testament to the impact of research in this space over the last 5-10 years. Although there is a lack of consensus about the \textit{best} deep learning reconstruction method, structured approaches to this inverse problem are likely to have greater adoption compared to purely data-driven approaches. In this respect, fully-convolutional image-to-image models that estimate attenuation correction from single- or multi-contrast MRI, trained with a supervised objective (e.g.~CT groundtruth) \textit{and} adversarial objectives (e.g.~real vs. fake), make the most compelling case. While these methods have achieved a high level of maturity and can improve quantitation of PET standardized uptake values, improvements to the reconstruction algorithm itself can offer more dramatic improvements. This is evidenced by the success of purely data-based deep learning methods in the literature. However, to build verifiable and trusted PET/MRI systems, a hybrid approach is still preferred amongst the community, as evidenced by the continued widespread use of MLAA and related variants. In this respect, newly introduced physics-driven methods that combine MLAA or joint-sparsity objectives with deep learning priors are the most promising since they also provide strong data consistency guarantees.

\vspace{-3mm}
\subsection{Opportunities for Innovation} There are numerous clinical applications that will benefit from enhanced PET/MRI, including improved early cancer detection and localization (e.g.~screening and post-surgery), improved measurement of response to therapy (e.g.~Amyloid PET for Alzheimer's), and dynamic imaging (e.g.~perfusion or cardiac imaging). Low-dose PET imaging is a particularly ripe area for innovation because it addresses the low-SNR in PET acquisition that is common to all the aforementioned applications. There is ample opportunity here for developing compressed sensing PET/MRI reconstruction techniques for both static and dynamic imaging, without resorting to purely data-driven approaches~\cite{ong2016beyond}. There is also a connection between autoencoding neural networks and compressed matrix factorization that has not been fully realized for PET/MRI, and could offer a bridge for more deep learning innovation. A major bottleneck for physics-driven PET/MRI reconstruction research seems to be the availability of open-format scanner specifications, which limits development of methods utilizing the anticipated system response of clinical scanners. This presents a challenge for methods seeking to improve the resolution of current PET/MRI, but this may be alleviated through increased adoption of open source tools, such as in~\cite{wadhwa2021pet}.

\vspace{-3mm}
\singlespacing
\renewcommand*{\bibfont}{\footnotesize}
\printbibliography
\end{document}